\documentstyle[12pt,aasms4]{article}
\begin{document}


\title{The Low Column Density Lyman-alpha Forest}
\author{Nickolay Y.\ Gnedin\altaffilmark{1,2} and Lam Hui\altaffilmark{1}}
\altaffiltext{1}{Department of Physics, Massachusetts Institute of Technology,
Cambridge, MA 02139; e-mail: \it gnedin@arcturus.mit.edu, lhui@space.mit.edu}
\altaffiltext{2}{Princeton University Observatory, Peyton Hall, 
Princeton, NJ 08544}


\load{\scriptsize}{\sc}

\def\ion#1#2{\rm #1\,\sc #2}
\def\HI{\ion{H}{i}}
\def\HII{\ion{H}{ii}}
\def\GI{\ion{He}{i}}
\def\GII{\ion{He}{ii}}
\def\GIII{\ion{He}{iii}}
\def\MH{{{\rm H}_2}}
\def\Hp{{{\rm H}_2^+}}
\def\Hm{{{\rm H}^-}}

\def\dim#1{\mbox{\,#1}}

\def\exp#1{\,e^{\displaystyle #1}\,}
\def\log{\mbox{\,ln}}

\def\figdir{.}
\def\placefig#1{}

\def\capSP{
The scatter plot of the column density of a hydrogen Lyman-alpha line
computed with the Stationary Phase approximation versus the exact
calculation. The Stationary Phase approximation estimates the column
density of an individual line to within a factor of 2.}

\def\capCM{
The column density distribution for the Standard CDM model with
$\sigma_8=0.7$ from the full calculation using the Zel'dovich approximation
and Miralda et al. (1996) line identification algorithm
({\it open circles\/}) and the Stationary Phase calculation using the
Zel'dovich approximation for the density distribution ({\it filled 
triangles\/}) versus approximate analytical treatment using the Stationary
Phase calculation and the lognormal model for the density field ({\it solid
lines\/}). The lower line corresponds to the same initial conditions as
the Zel'dovich approximation (the
same smoothing scale), and the upper line corresponds to the smaller
smoothing scale that gives the same rms density fluctuation as the
Zel'dovich approximation. The number of lines drops drastically for the 
simulated distributions at low column densities due to finite 
resolution.}

\def\capLL{
Comparison between six cosmological models and observations ({\it solid
squares\/}) at $z=3$. The shaded region shows the predicted range for the
models with both $T_{\rm gas}$ and $T_J$ taking values between 
$3\times10^3\dim{K}$ and $3\times10^4\dim{K}$.}

\def\capBF{
Comparison between three best-fit cosmological models and observations 
for two currently favorable values of the cosmological baryon density.}

\def\tabMM{
\begin{table}
\caption{Cosmological Models\label{tabMM}}
\medskip
$$
\begin{tabular}{lllllll}
Model & $\Omega_0$ & $\Omega_\Lambda$ & $\Omega_\nu$ & $h$ & $n$ & $\sigma_8$ 
							\\ \tableline
SCDM1 & $1$        & $0$              & $0$      & $0.50$ & $1$    & $0.40$ \\
SCDM2 & $1$        & $0$              & $0$      & $0.50$ & $1$    & $1.15$  \\
LCDM1 & $0.35$     & $0.65$           & $0$      & $0.70$ & $0.96$ & $0.67$  \\
LCDM2 & $0.35$     & $0.65$           & $0$      & $0.70$ & $1$    & $0.85$  \\
LCDM3 & $0.40$     & $0.60$           & $0$      & $0.65$ & $1$    & $0.84$  \\
CHDM1 & $1$        & $0$              & $0.2$    & $0.50$ & $1$    & $0.76$  \\
\end{tabular}
$$
\end{table}
}

\def\tabPP{
\begin{table}
\caption{Spectral Parameters at $T_J=10^4\dim{K}$\label{tabPP}}
\medskip
$$
\begin{tabular}{lllllll}
Model & $\sigma_0$ & $\gamma$ & $R_* (h^{-1}\dim{Mpc})$	\\ \tableline
SCDM1 & $1.01$     & $0.52$   & $0.105$ \\
SCDM2 & $2.90$     & $0.52$   & $0.105$ \\
LCDM1 & $1.11$     & $0.51$   & $0.180$ \\
LCDM2 & $1.50$     & $0.52$   & $0.178$ \\
LCDM3 & $1.51$     & $0.52$   & $0.167$ \\
CHDM1 & $1.18$     & $0.49$   & $0.106$ \\
\end{tabular}
$$
\end{table}
}

\begin{abstract}

We develop an {\it analytical\/} method based on the lognormal
approximation to compute the column density
distribution of the Lyman-alpha forest in the low column density limit.
We compute the column density
distributions for six different cosmological models and found that
the standard, {\it COBE\/}-normalized CDM model cannot fit
the observations of the Lyman-alpha forest at $z=3$. The amplitude
of the fluctuations in that model has to be lowered by a factor of 
almost 3 to match observations. However, the currently
viable cosmological models like the slightly tilted {\it COBE\/}-normalized 
CDM+$\Lambda$ model, the CHDM model with $\Omega_\nu=0.2$, and the
low-amplitude Standard CDM model are all in agreement with
observations,
to within the accuracy of our approximation,
for the value of the cosmological baryon density at or higher than
the old Standard Big Bang nucleosynthesis value of $\Omega_bh^2=0.0125$ for
the currently favored value of the ionizing radiation intensity.
With the low value for the baryon density inferred by Hogan \& 
Rugers\markcite{HR96} (1996), the models can only marginally match
observations.

\end{abstract}


\section{Introduction}

The Lyman-alpha forest, numerous weak hydrogen absorption lines in the 
spectra of distant quasars, have been the focus of extensive study for 
more that two decades. Recent cosmological hydrodynamic simulations 
greatly advance our understanding of it and allow 
unambiguous confrontation between observations with theoretical models  
(Cen et al.\markcite{CMOR94} 1994;
Miralda-Escude et al.\markcite{MCOR95} 1995; Hernquist et al.\markcite{HKWM96}
1996; Zhang et al.\markcite{ZMAN96} 1996).

Numerical simulations show that the Lyman-alpha forest forms as a
result of absorption of the quasar light by neutral hydrogen in
inhomogeneities on small scales very much like the 
Lyman limit systems, arising from absorption by galaxies on larger scales.
One of the predictions of such models is that the low column
density Lyman-alpha forest ($N_{\HI}\la10^{14}\dim{cm}^{-2}$)
resides in regions of low overdensity and even underdense
regions of the universe. While higher column density Lyman-alpha
systems\footnote{We intentionally avoid using the term ``clouds'' since
the forest mostly consists of fluctuations in the intergalactic
medium rather than discrete objects.}
form in significantly overdense regions ($\delta\equiv\delta\rho/\bar\rho
\ga3-10$), which 
requires  the use of numerical simulations to compute their properties
accurately, the low column density systems may be studied 
analytically since there exist reliable approximations to study 
structure formation in the regime of low overdensity. The Lyman-alpha
forest arising from these slightly overdense and underdense regions
is the subject of this {\it Letter\/}.

\section{Column Density from the Stationary Phase Integration Around
the Peak}

We first start with deriving an approximate expression for the column
density of an absorption line arising from a density peak in velocity space.
Let $\tau_0$ be the Lyman-alpha optical depth for a homogeneous medium at
the average cosmological density at redshift $z$ along a line-of-sight to a 
distant quasar (see Jenkins \&  
Ostriker\markcite{JO91} 1991).
We will consider the range of $z\sim3$ when 
hydrogen was in ionization equilibrium with the radiation field
which had photoionization rate $\Gamma\equiv4.3\times10^{-12}J_{-21}
\dim{s}^{-1}$. $J_{-21}$, which characterizes the radiation intensity, is 
assumed to be $0.5$. Since the number
density of neutral hydrogen is proportional to the square of the density
times the recombination coefficient,
we can now compute the column density $N_{\HI}$
of a Lyman-alpha absorption line:
\begin{equation}
	N_{\HI}\sigma_{\rm Ly-\alpha} = {\dot{a}\tau_0\over c}
	\int \exp{2(1-\alpha)\xi} d\,x
	\label{colden}
\end{equation}
where $\xi\equiv\log(1+\delta)$, $\delta$ being the overdensity, and the 
integral is taken over the  
spatial comoving coordinate $x$ in the vicinity of the density peak.
In this expression we implicitly assume that each Lyman-alpha line
is dominated by a single density peak and we ignore peculiar velocity effects. 
The exact limits of 
integration are assumed to be unimportant as long as the dominant peak 
is included.
The parameter $\alpha$ measures the deviation
from the isothermality; if the recombination coefficient changes with the
temperature as $T^{-\beta}$, and the equation of state is $T\sim\rho^\gamma$,
then $\Omega_{\HI} \sim \rho^2 T^{-\beta} \sim \rho^{2-\beta\gamma}$ and
$\alpha=\beta\gamma/2$. For typical values of $\beta=0.7$ and $\gamma=0.5$
(as shown in Hui \& Gnedin\markcite{HGB96} 1996),
the correction for non-isothermality is only $\beta\gamma/2\la0.2$. 
In the following, we will assume $\alpha=0$, but will allow for a range
of temperatures, from $3,000\dim{K}$ to $30,000\dim{K}$, at every value
of density, to account
for not only power-law equations of state, but also other
relationships between the density and the temperature of the gas. However,
we will keep $\alpha$ in our calculations for the sake of generality.

The integral in (\ref{colden}) can be computed using the 
Stationary Phase method,
\begin{equation}
	N_{\HI}\sigma_{\rm Ly-\alpha} = {\dot{a}\tau_0\over c}
	\exp{2(1-\alpha)\xi_c}\sqrt{\pi\over -(1-\alpha)\xi^{\prime\prime}_c},
	\label{nfin}
\end{equation}
where $\xi^{\prime\prime}$ is the second derivative of the logarithm
of the density with respect to the comoving coordinate along the line-of-sight
and the subscript $c$ denotes evaluation at the density peak.

In order to proceed further and compute the column density distribution,
we need to know the distribution function of the density and its derivatives.
In the linear regime, the distribution function is gaussian, and the
calculation can be performed fully analytically. However, we want to 
apply the approximation (\ref{nfin}) to a wider range of overdensities,
particularly for underdense regions. We therefore use the lognormal
distribution for the density (Bi, Borner, \& Chu\markcite{BBC92} 1992; 
Coles, Melott, \& Shandarin\markcite{CMS93} 1993), 
which is mainly based upon the assumption 
that the velocity distribution remains gaussian for larger values of
overdensities than the density distribution does. Since the continuity equation
in the expanding coordinates can be written as:
\begin{equation}
	{d\xi\over d\,t} = -{1\over a}{\partial v^i\over\partial x^i},
\end{equation}
where $d/d\,t$ denotes the Lagrangian derivative and $v$ is the
peculiar velocity, we can see that
the density distribution stays close to lognormal (i.e.\ $\xi$ is
normally distributed) as long as $v$ is gaussian.

Since a one-dimensional slice of a three-dimensional gaussian random field
is also a gaussian random field, we can use the BBKS formalism
(Bardeen et al.\markcite{BBKS86} 1986) to derive the column density
distribution of the low column density Lyman-alpha forest. 
Let us introduce the quantity $g$ as follows:
\begin{equation}
	g \equiv \log\left(N_{\HI}\sigma_{\rm Ly-\alpha}c\over\tau_0
	\sqrt{\pi}\dot{a}R_*\right) + \sigma_0^2(1-\alpha) = 
	2(1-\alpha)(\xi - \bar{\xi}) - {1\over2}\log\left[-R_*^2
	(1-\alpha)\xi^{\prime\prime}\right],
	\label{gdef}
\end{equation}
where the parameters $\sigma_0$, $R_*$, and $\gamma$ (which appears below)
are defined as in the BBKS paper, and $\bar{\xi}=-\sigma_0^2/2$ is the
average value of $\xi$.

Then the column density distribution (number of absorption 
lines per unit column 
density per unit redshift) of the Lyman-alpha forest is given
by the following simple expression:
\begin{equation}
	{d^2{\cal N}_{\rm Ly-\alpha}\over dN_{\HI}\,dz} = 
	{c\over N_{\HI}H_0\sqrt{\Omega_0(1+z)^3}}
	{dn_{\rm pk}\over dg},
	\label{ndist}
\end{equation}
where $dn_{\rm pk}/dg$ is the one-dimensional comoving number density of 
peaks per unit interval of
$g$, which can be easily computed using the BBKS methodology, and is given
by the following expression:
\begin{equation}
	{dn_{\rm pk}\over dg} = 
	{1\over2(2\pi)^{3/2}R_*\sigma_0^3\gamma^2\sqrt{9/5-\gamma^2}(1-\alpha)}
	\int_0^\infty x\exp{-Q(g,x)}\,dx,
	\label{dint}
\end{equation}
where
\begin{equation}
	Q(g,x) = {1\over2\sigma_0^2(9/5-\gamma^2)}\left({9\over5}\Delta^2-
	2x\Delta+{1\over\gamma^2}x^2\right),
\end{equation}
and
\begin{equation}
	\Delta(g,x) \equiv \xi-\bar\xi = 
	{1\over 2(1-\alpha)}\left(g+{1\over2}\log((1-\alpha)x)\right).
\end{equation}
The integral in (\ref{dint}) can be easily computed numerically
for a given values of $\sigma_0$ and $\gamma$.

In order to test our approximation, we compute the exact column densities
using the Zel'dovich approximation and Miralda et al.\markcite{MCOR96} (1996)
line
identification algorithm and for each identified line we apply the
Stationary Phase approximation. We find that the Stationary
Phase column density is within a factor of two from and somewhat below 
(for higher column densities) the exact column density.

\begin{figure}
\epsscale{0.55}
\plotone{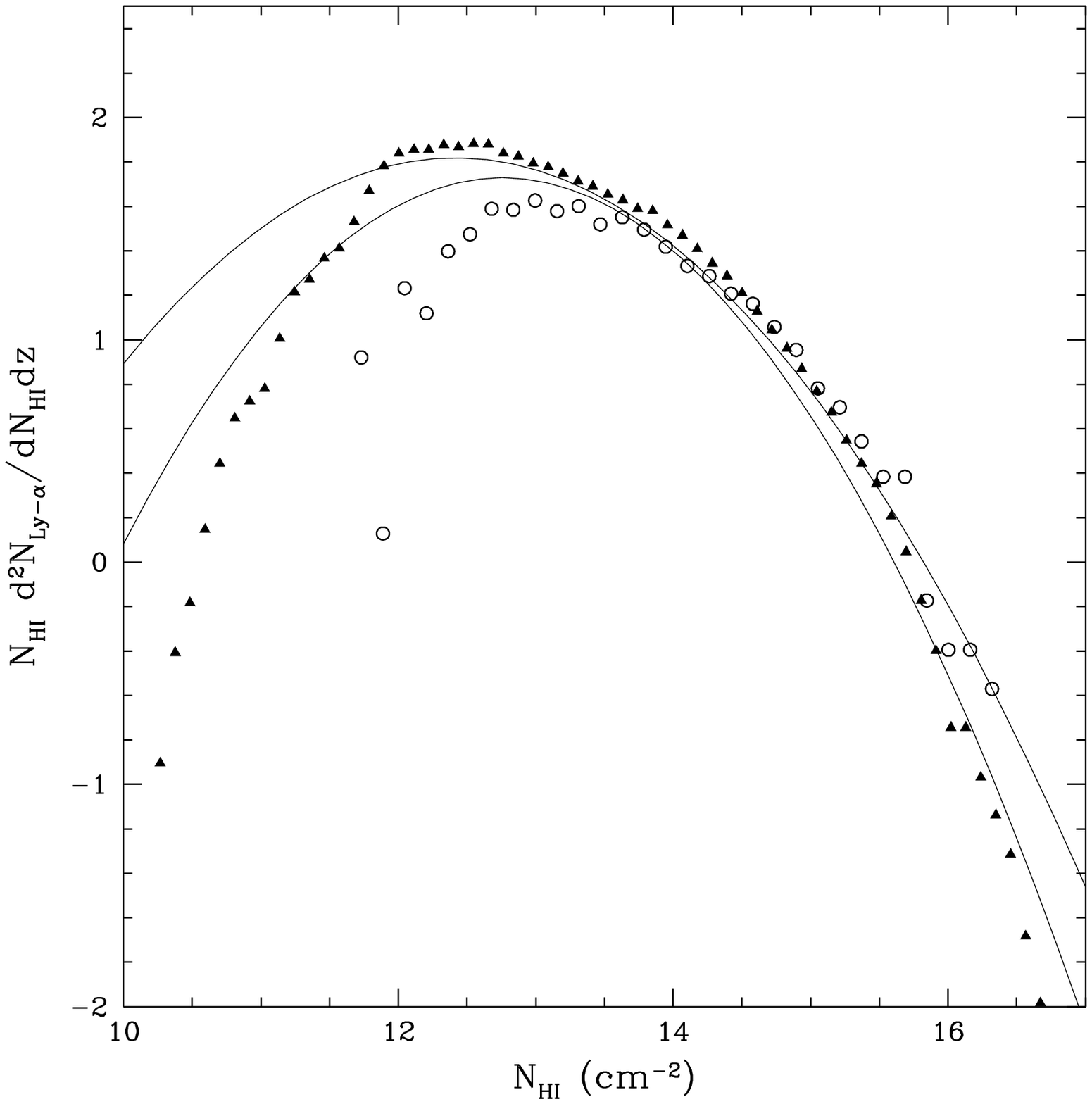}
\caption{\label{figCM}\capCM}
\end{figure}

Since our approximation assumes the lognormal
density distribution, we test it against the Zel'dovich approximation, 
as shown in 
Fig.\ref{figCM}. We show with open symbols the exact column density 
distribution (multiplied by the column density to better demonstrate the
differences) from our Zel'dovich calculation (see Hui \& Gnedin\markcite{HGB96}
1996 for details) together with the Stationary Phase calculation using the
density distribution from the Zel'dovich approximation (filled triangles).
\footnote{The exact computation lies below the Stationary Phase +
Zel'dovich approximation at low column density because the Miralda
et al.\markcite{MCOR96} (1996) line identification algorithm significantly
underestimates the number of low column density lines (M.\ L.\ Norman,
private communication).} The number of absorption lines drops drastically for
$N_{\HI}<10^{12}\dim{cm}^{-2}$ due to the finite resolution of our
simulations.
The two solid lines show the approximate analytical calculation for two
choices of the smoothing scale: the lower curve has the same smoothing as
the Zel'dovich approximation, and the upper curve has a smaller smoothing
scale so as to have the same rms density fluctuation as the Zel'dovich
approximation. In general, we again see that the analytical approximation
gives a good guess for the real column density distribution, perhaps,
slightly underestimating the column density distribution.

\section{Testing the Models}

We consider six different models whose parameters are compiled in Table
\ref{tabMM}. Our treatment only requires three spectral parameters,
$\sigma_0$, $\gamma$, and $R_*$. However, for a power spectrum
which behaves like $k^{-3}$ at large $k$, integrals for $\gamma$ and
$R_*$ diverge. The power spectrum, therefore, has to be cut off at
some scale. The natural cut-off scale would be the Jeans scale but
the power spectrum at scales smaller than it behaves
like $k^{-7}$ (Bi et al.\markcite{BBC92} 1992), 
and integrals needed to compute $R_*$ and $\gamma$ diverge.
We, therefore, apply the exponential cut-off at the Jeans scale,
\begin{equation}
	P_{\rm gas}(k) = P_{\rm DM}(k)\exp{-(k/k_J)^2}\ 
	\label{cutoff}
\end{equation}
where
\begin{equation}
	k_J \equiv 7.4 (\Omega_0(1+z)10^4\dim{K}/T_J)^{1/2} h\dim{Mpc}^{-1},
	\label{jeansk}
\end{equation}
but we treat the temperature $T_J$ at which we compute the Jeans scale
as an independent parameter not necessarily equal to the gas temperature
$T_{\rm gas}$. We note here that this cut-off of the
power spectrum is equivalent to smoothing the one-dimensional density
distribution on the scale of Doppler broadening. Even though Doppler
broadening does not affect the value of the column density, the 
identification of
a line depends on the value of $b$ ($b = \sqrt{2 k_B T/ m_p}$ where
$m_p$ is the proton mass) as two narrow peaks that are separated by
a distance much less than $b$ in velocity space would not be counted as
two lines but rather as one line. Since this procedure is somewhat
dependent on the line identification algorithm, we use the freedom
in $T_J$ as a way to account for differences between different line
identification algorithms. We then consider the range of temperatures 
from $3\times10^3\dim{K}$ to $3\times10^4\dim{K}$ for both
$T_J$ and the gas temperature $T_{\rm gas}$ in our calculations and we put
$\alpha=0$. The range of temperatures
at a given density should be enough to cover possible variations 
for any other reasonable equation of state.  
We assume $J_{-21}=0.5$ throughout.


\begin{table}
\caption{Cosmological Models\label{tabMM}}
\medskip
$$
\begin{tabular}{lllllll}
Model & $\Omega_0$ & $\Omega_\Lambda$ & $\Omega_\nu$ & $h$ & $n$ & $\sigma_8$ 
							\\ \tableline
SCDM1 & $1$        & $0$              & $0$      & $0.50$ & $1$    & $0.40$ \\
SCDM2 & $1$        & $0$              & $0$      & $0.50$ & $1$    & $1.15$  \\
LCDM1 & $0.35$     & $0.65$           & $0$      & $0.70$ & $0.96$ & $0.67$  \\
LCDM2 & $0.35$     & $0.65$           & $0$      & $0.70$ & $1$    & $0.85$  \\
LCDM3 & $0.40$     & $0.60$           & $0$      & $0.65$ & $1$    & $0.84$  \\
CHDM1 & $1$        & $0$              & $0.2$    & $0.50$ & $1$    & $0.76$  \\
\end{tabular}
$$
\end{table}

The first two models we consider are standard CDM models, one is
with $\sigma_8=0.4$
at $z=0$ and the other is {\it COBE\/}-normalized. We use the BBKS transfer
function to compute spectral parameters $\sigma_0$, $\gamma$, and $R_*$
for this model. 

The next three models are the {\it COBE\/}-normalized 
CDM+$\Lambda$ models whose
transfer functions cannot be fit by the BBKS formula accurately. We therefore
compute transfer functions for LCDM models using the linear gravity
code similar to (but different from) COSMICS package 
(Bertschinger\markcite{B95} 1995) and 
we use exact transfer functions to compute
the spectral parameters. The first LCDM model assumes a small tilt
in the primordial power spectrum and 25\% of gravity waves added
in quadrature as in Kofman, Gnedin, \& Bahcall\markcite{KGB93} (1993).


\begin{table}
\caption{Spectral Parameters at $T_J=10^4\dim{K}$\label{tabPP}}
\medskip
$$
\begin{tabular}{lllllll}
Model & $\sigma_0$ & $\gamma$ & $R_* (h^{-1}\dim{Mpc})$	\\ \tableline
SCDM1 & $1.01$     & $0.52$   & $0.105$ \\
SCDM2 & $2.90$     & $0.52$   & $0.105$ \\
LCDM1 & $1.11$     & $0.51$   & $0.180$ \\
LCDM2 & $1.50$     & $0.52$   & $0.178$ \\
LCDM3 & $1.51$     & $0.52$   & $0.167$ \\
CHDM1 & $1.18$     & $0.49$   & $0.106$ \\
\end{tabular}
$$
\end{table}

Finally, we study the {\it COBE\/}-normalized CHDM model with currently
favored value of $\Omega_\nu=0.2$ and no tilt. 
We use the Ma\markcite{M96} (1996) 
transfer function in our calculation. Table \ref{tabPP} lists
spectral parameters for the six models at $z=3$ and $T=10^4\dim{K}$.

\begin{figure}
\epsscale{1.15}
\plotone{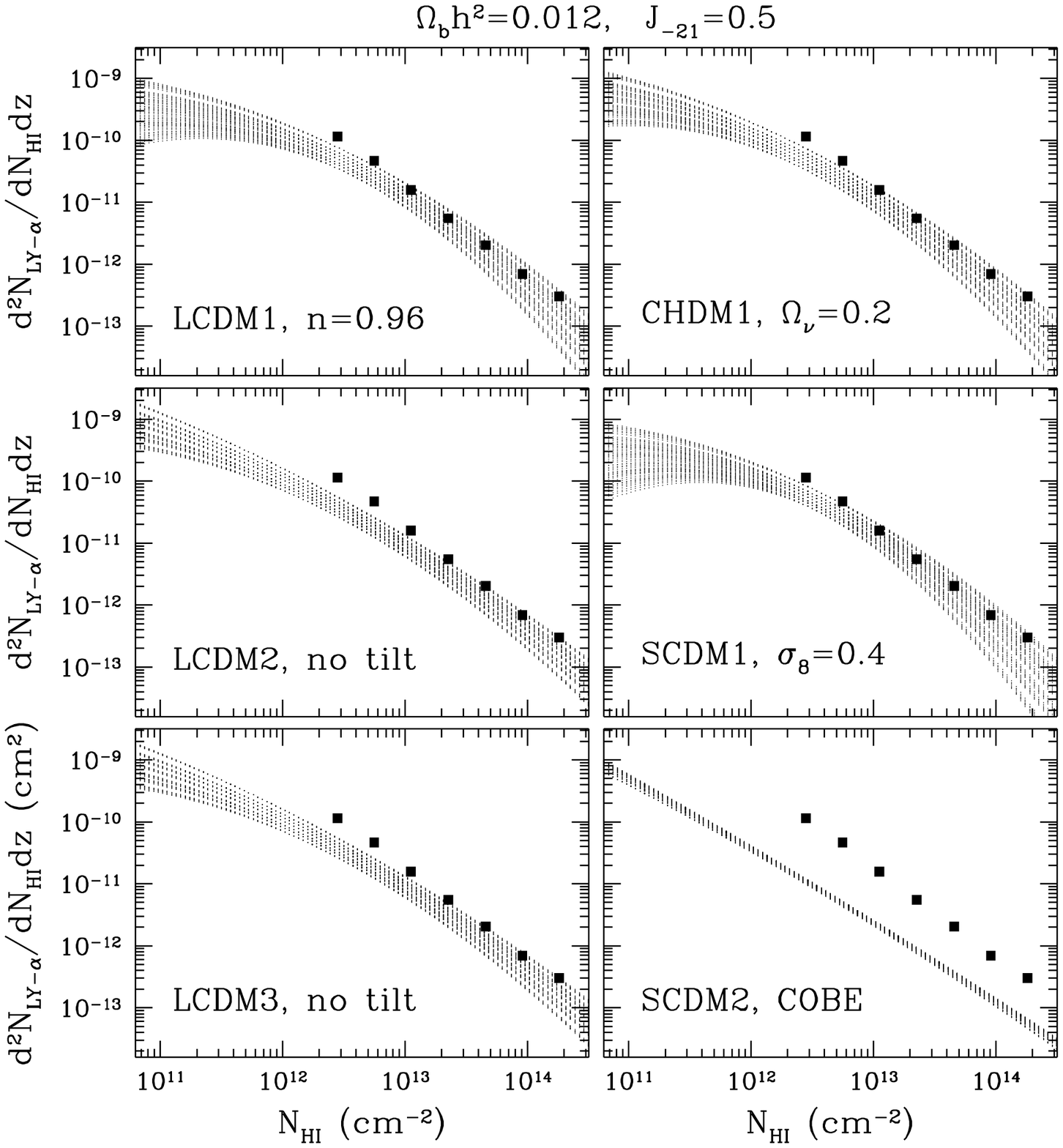}
\caption{\label{figLL}\capLL}
\end{figure}

Fig.\ref{figLL} shows the computed column density distributions for the
above six models where both $T_J$ and $T_{\rm gas}$ independently
take values from $3\times10^3\dim{K}$ to $3\times10^4\dim{K}$ at $z=3$. 
We use the old Standard Big Bang Nucleosynthesis  value of
$\Omega_bh^2=0.0125$ for the baryon density 
in these calculations.
We emphasize again that since the
temperature of the intergalactic gas in not known precisely, and
it may also depend on local conditions, it is appropriate to consider 
the hatched area
as the region where a more
accurate calculation would lie. The width of the hatched area also
corresponds to the accuracy of the analytical approximation as can
be judged from Fig.\ref{figCM}.
We also show the
observed abundances from Hu et al.\markcite{HKCS95} (1995) with
filled squares. The size of the squares roughly corresponds to the 
observational uncertainties.

As could be expected, the {\it COBE\/}-normalized standard CDM does
not fit the data, but the low-amplitude SCDM, slightly tilted LCDM and
the CHDM models provide a good fit to the data
for low column densities. At higher column densities,
$N_{\HI}\ga10^{14}\dim{cm}^{-2}$, our approximation breaks down since
those column densities correspond to significantly overdense regions.
The two no-tilt LCDM models produce slightly too few low
column density Lyman-alpha systems because their $\sigma_0$ is too
high and their underdense regions are too underdense; they may be
only marginally consistent with observations.

It is remarkable that the column density distribution is by
far most sensitive to the value of $\sigma_0$. It is the value
of $\sigma_0(T_J=10^4\dim{K})\approx1.1$ 
that ir responsible to the match with observations for
SCDM1, LCDM1, and CHDM1 models. We found
that the dependence of the results on other parameters ($\Omega_0$,
$h$, $J_{-21}$, $T$, $\gamma$, and $R_*$) is weaker than dependence on 
$\sigma_0$, and for
any set of those six parameters (within reasonable limits)
it is possible to adjust the value of $\sigma_0$ to give a good fit to
observations.

\begin{figure}
\epsscale{1.15}
\plotone{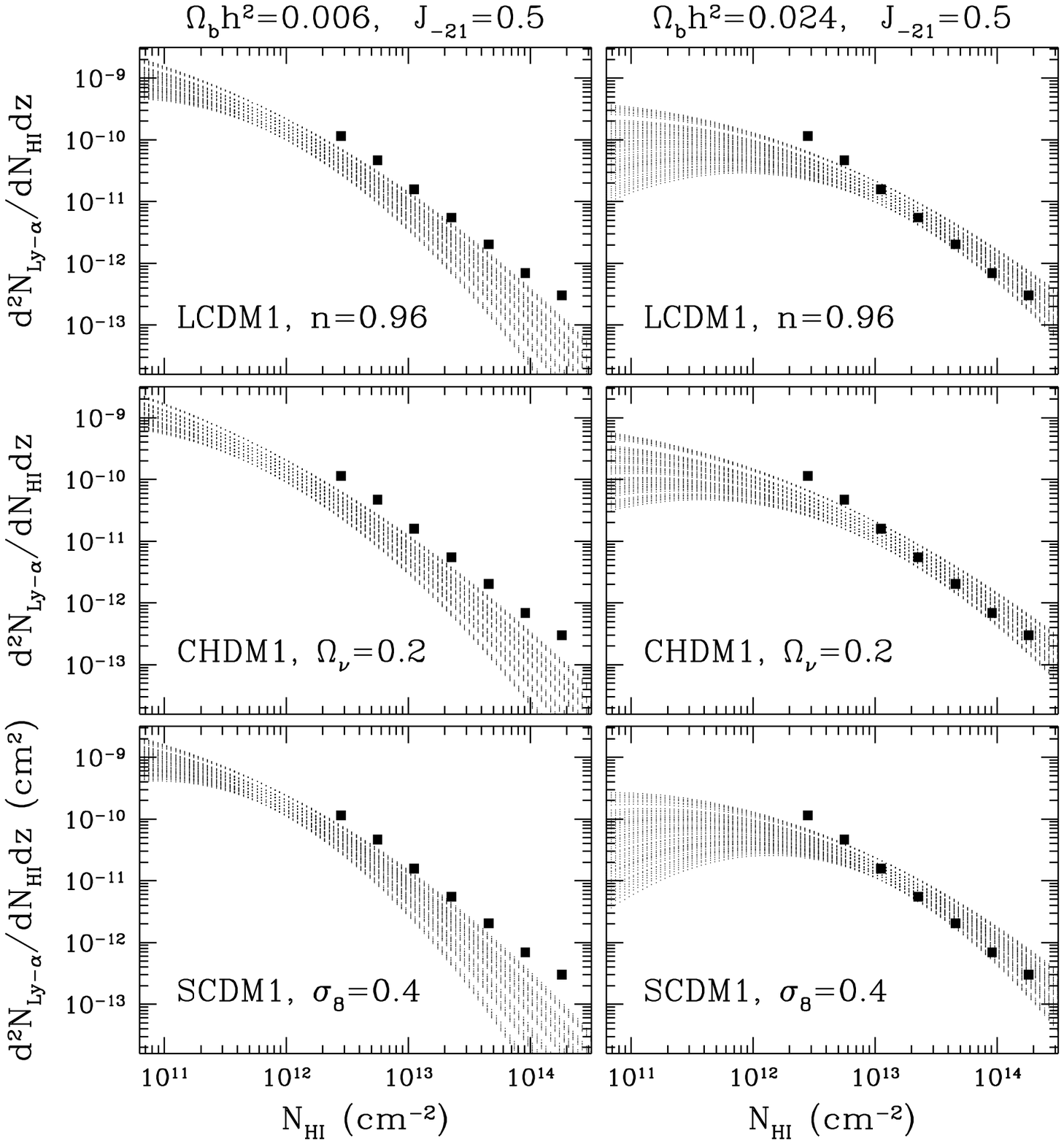}
\caption{\label{figBF}\capBF}
\end{figure}

Finally, we address the question of uncertainty in the baryon density.
Recently there have been new determinations of the cosmological
baryon density using the high-redshift deuterium measurements giving
two contradicting values of $\Omega_bh^2=0.006$ (Hogan \& 
Rugers\markcite{HR96} 1996) and $\Omega_bh^2=0.024$ (Tytler \&
Burles\markcite{TB96} 1996). We plot the results for the three best-fit
models, LCDM1, CHDM1, and SCDM1, for those two values of the baryon
density and $J_{-21}=0.5$ in Fig.\ref{figBF}. We note that the high value for
the baryon density improves the agreement between the models and the data,
while with the low value for the baryon density, the models are only 
marginally consistent
with observations.

\section{Conclusions}

We have developed an {\it analytical\/} method to compute the column density
distribution for the Lyman-alpha forest in the low column density limit,
$N_{\HI}\la10^{14}\dim{cm}^{-2}$. We computed the column density
distributions for six different cosmological models and found that
the standard, {\it COBE\/}-normalized CDM model cannot fit
the observations of the Lyman-alpha forest at $z=3$. The amplitude
of density fluctuations in that model has to be lowered by a factor of 
almost 3 to match observations. The {\it COBE\/}-normalized 
CDM+$\Lambda$ model with a slight tilt, the {\it COBE\/}-normalized CHDM 
model with $\Omega_\nu=0.2$, and the low-amplitude Standard CDM model
are all in agreement with observations for a range of values for
the cosmological baryon density and for the currently favored value
of the ionizing background radiation intensity of $J_{-21}=0.5$.
The no-tilt LCDM models without gravity
wave contribution produces slightly too few low column density Lyman-alpha
systems. 
We also note that since at high redshift, $z\ga2$, 
all cosmological models have similar rate of growth
of perturbations on scales of interest (including the CHDM model, since the
neutrino free-streaming scale is larger than the characteristic scale for the
low column density Lyman-alpha forest), and
since the three models, LCDM1, CHDM1, and SCDM1, have similar predictions
at $z=3$, there will be little differences in their predictions at $z=2$
or $z=4$. 
We conclude that, based on the lognormal approximation, the column density 
distribution alone provides only a weak constraint on currently viable
cosmological models unless the physical state of the intergalactic medium 
is known more precisely from independent observations.
A more accurate approximation would also be useful in decreasing the width 
of the hatched area in Fig.\ref{figLL}, which 
takes into account our ignorance of the conditions  
of the intergalactic medium as well as the accuracy of the 
lognormal approximation.

The computed column density distribution is by far most sensitive to
the amplitude of fluctuations at the Jeans scale, $\sigma_0$.
We conclude that models with $\sigma_0 \sim 1$ (at $T_J=10^4\dim{K}$
and $z=3$) are capable of reproducing the observational data for a
large range of other parameters.

Finally, we note that with the low value for baryon density as inferred by
Hogan \& Rugers\markcite{HR96} (1996), all of the models tested 
here are only marginally consistent with observations.

\acknowledgements

The authors thank Edmund  Bertschinger for
numerous fruitful discussions and valuable suggestions. We are also grateful
to Sergei Shandarin for very helpful comments.
This work was supported in part by the NSF grant AST-9318185.

%
%
%
%
%
%
%
%
%

\end{document}